\documentclass[prb,twocolumn,bibnotes,showpacs,amsbsy,floatfix,amsbsy,floatfix,superscriptaddress]{revtex4-1}
\usepackage{epsfig,color}
\usepackage[cp1250]{inputenc}
\usepackage{amsmath}
\usepackage{amsfonts}
\usepackage{color}
\usepackage{wasysym}
\usepackage{latexsym} 
\usepackage{float}

\begin{document}

\title{Chiral and Non-Chiral Edge States in Quantum Hall  Systems\\ with Charge Density Modulation}
\author{Pawe\l{} Szumniak}
\affiliation{Department of Physics, University of Basel, Klingelbergstrasse 82, 4056 Basel, Switzerland}
\affiliation{AGH University of Science and Technology, Faculty of
Physics and Applied Computer Science,\\
al. Mickiewicza 30, 30-059 Krak\'ow, Poland}
\author{Jelena Klinovaja}
\affiliation{Department of Physics, University of Basel, Klingelbergstrasse 82, 4056 Basel, Switzerland}
\author{Daniel Loss}
\affiliation{Department of Physics, University of Basel, Klingelbergstrasse 82, 4056 Basel, Switzerland}
\date{\today}



\begin{abstract}
We consider a  system of weakly coupled wires with quantum Hall effect (QHE) and in the presence of a spatially periodic modulation of the chemical potential along the wire, equivalent to a charge density wave (CDW). We investigate the competition between the two effects which both open a gap.
We show that by changing the ratio between the amplitudes of the CDW modulation and the tunneling between wires, one can switch between non-topological CDW-dominated phase to topological  QHE-dominated phase. 
Both phases host edge states of chiral and non-chiral nature robust to on-site disorder. However, only in the topological phase, the edge states are immune to disorder in the phase shifts of the CDWs. We provide analytical solutions for filling factor $\nu=1$ and study numerically effects of disorder as well as present numerical results for higher filling factors.
\end{abstract}

\maketitle
{\it Introduction.} 
Over the last decades topological states of matter  attracted a lot of attention both theoretically and experimentally. 
The striking stability of the quantum Hall effect (QHE)~\cite{PhysRevLett.45.494,PhysRevLett.48.1559} 
can be linked to  topology~\cite{PhysRevLett.49.405}. The time-reversal invariant cousins of the QHE are the two-dimensional (2D) topological insulators (TIs) for which many candidate materials were found or synthesized in recent years~\cite{RevModPhys.82.3045}.
However, despite great progress, the conductance quantization in TI materials is still not as perfect as in the QHE. 
Thus, the experimental focus has shifted in recent years to a more direct study of the edge state physics in TIs, for instance via Fraunhofer patterns~\cite{hart2014induced, pribiag2015edge} and SQUID probes \cite{PhysRevLett.113.026804,Wang28082015}. However, the edge states, especially in clean samples, could also be of non-topological origin, for example, due to Tamm-Shockley states~\cite{ITamm, PhysRev.56.317,PhysRevLett.108.136803}. 

Recently, edge state behavior was observed in 2D GaSb heterostructures, but in a regime that is believed to be non-topological, and thus 
challenging the standard interpretation of this system as a TI \cite{Nichele2015}. The origin of this unexpected observation is still unclear but it raises the intriguing question whether edge states could not occur in both phases, in the topological as well as in the trivial one, but with different signatures such as e.g. being helical (chiral) in one phase vs. non-helical (non-chiral) in the other. 
In other words, the system could host topological edge states for one set of parameters while there exist non-topological edge states for an other one.
It is thus of fundamental interest to see if  realistic models can be constructed which demonstrate that, in principle, these two scenarios 
do not need to exclude each other.

\begin{figure}[bt!]
\centering
\includegraphics[width=8.6cm]{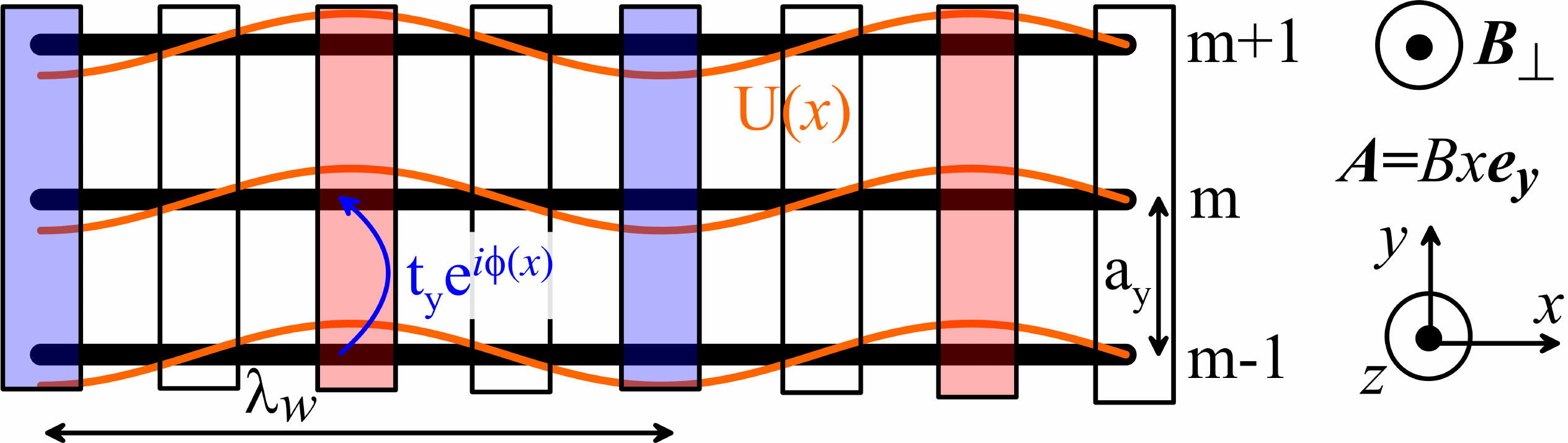}
\caption{A 2D strip of weakly coupled wires in a perpendicular magnetic field $B_\perp$ applied along the $z$ axis. The wires (thick black lines) aligned along the $x$ axis and labeled by the index $m$ are weakly coupled with tunnel amplitude $t_y$ in the $y$ direction. The vector potential ${\bf A}=B x \boldsymbol{e}_y$ is chosen to be in the $y$ direction such that the phase $\phi (x)=eB a_y x / \hbar c $ acquired in the tunneling  is position dependent, where $a_y$ is
the distance between neighboring wires. There is a CDW modulation $U(x)$ (orange wavy line) of the chemical potential inside the wires, with period
$\lambda_w$ tuned to the Fermi wave length $\lambda_F$.
}
  \label{setup}
\end{figure}

In the present work, we propose a system related to the QHE regime where exactly such a mixed behavior of edge state physics can emerge. The system we consider is given by a 2D array of tunnel coupled wires in the presence of a magnetic field  and  charge density waves (CDWs) inside the wires. This provides two different mechanisms (QHE and CDW) for inducing gaps and edge states which can compete with each other.
Such CDWs may be induced intrinsically by electron-electron interactions \cite{giamarchi2003quantum, FQHE_CDW_Jelena, 1994lanm.proc...15B}, extrinsically by  periodically arranged gates inducing spatial modulations of the chemical potential, or by an internal superlattice structure \cite{algra2008twinning, PhysRevLett.86.1857}, see Fig.~\ref{setup}.   We show that by tuning the ratio between the amplitude of the CDW modulation and the tunneling amplitude between the wires the system undergoes a phase transition between a non-topological (CDW dominated) phase, which supports predominantly non-chiral edge states, and a topological (QHE dominated) phase, which supports predominantly  chiral edge states.  However, in both phases, one can find both chiral and non-chiral regimes.
These results are supported by both numerical and analytical calculations.
We confirm numerically that, as expected, the topological chiral states are less susceptible to disorder.

{\it Tight-binding model.} 
We consider a 2D  coupled wire construction~\cite{lebed1986anisotropy, PhysRevB.43.11353, PhysRevLett.111.196401, PhysRevB.89.085101, PhysRevB.90.115426, PhysRevB.89.104523, PhysRevB.90.201102, PhysRevB.90.235425, PhysRevB.91.085426, PhysRevB.91.241106} in the presence of a perpendicular uniform magnetic field, see Fig. \ref{setup}. 
We assume that the propagation is anisotropic in the $xy$ plane, mainly for analytical convenience.
The tunneling amplitude along the wires, aligned in the $x$ direction, is thus larger than the one between the wires in the $y$ direction. This allows us to treat the wires as independent one-dimensional channels only weakly coupled to their neighboring wires. In addition, we include a CDW modulation along the  wire.
The system is then described by the following tight-binding Hamiltonian,
\begin{align}
 &H=\sum_{n,m}\Big( -t  c^{\dag}_{n+1,m}c_{n,m} - t_y e^{in\phi}  c^{\dag}_{n,m+1}c_{n,m}   \\ 
&- [U_0\cos(2k_{w}na_x+\varphi)+\mu/2] c^{\dag}_{n,m}c_{n,m} + H.c.\Big) \nonumber,
\label{eq:H_2D}
\end{align}
where $c_{n,m}$ is the annihilation operator acting on the electron at a site $(n,m)$ of the lattice with the lattice constant $a_x$ ($a_y$) in the $x$ ($y$) direction. For simplicity we consider  spinless electrons in this work. We choose the hopping amplitude along the $x$ direction $t>0$ to be much larger than the hopping along the $y$ direction $t_y>0$. The uniform magnetic field applied in the $z$ direction, $\boldsymbol{B}=B\boldsymbol{e}_z$, and the corresponding vector potential $\boldsymbol{A}=Bx\boldsymbol{e}_y$ is chosen along the $y$ axis,
yielding the orbital Peierls phase $\phi=eBa_xa_y/\hbar c$. The chemical potential $\mu$  is modulated with the CDW amplitude $2U_0>0$ and the period $\lambda_{w}=\pi/k_{w}$. The angle $\varphi$ is the phase of the CDW at the left edge of the wire ($n=0$).

With this choice of the vector potential $\bf A$, the system is translation invariant in the $y$ direction, thus, we can introduce the momentum $k_y$ via Fourier transformation $c_{n,m}=\frac{1}{\sqrt{N_y}}\sum_{k_y}c_{n,k_y}e^{-imk_y a_y}$, where $N_y$ is the number of lattice sites in the $y$ direction. The Hamiltonian becomes diagonal in $k_y$ space,
\begin{align}
&H=\sum_{n,k_y}\Big ( [-t  c^{\dag}_{n+1,k_y} c_{n,k_y} +H.c. ] - c^{\dag}_{n,k_y}c_{n,k_y} \\
&\times [\mu + 2 U_0\cos(2k_{w}na_x+\varphi) + 2 t_y \cos(n\phi+k_ya_y)] \Big ) \nonumber
\label{eq:H_1D_ky}
\end{align}
As a result, the eigenfunctions of $H$ factorize as $e^{i k_y y }\psi_{k_y}(x)$, with $x=na_x$, $y=ma_y$. From now on, we focus on $\psi_{k_y}(x)$ and treat $k_y$ as a parameter.\\
\begin{figure*}[ht!]
\centering
\includegraphics[width=18.3cm]{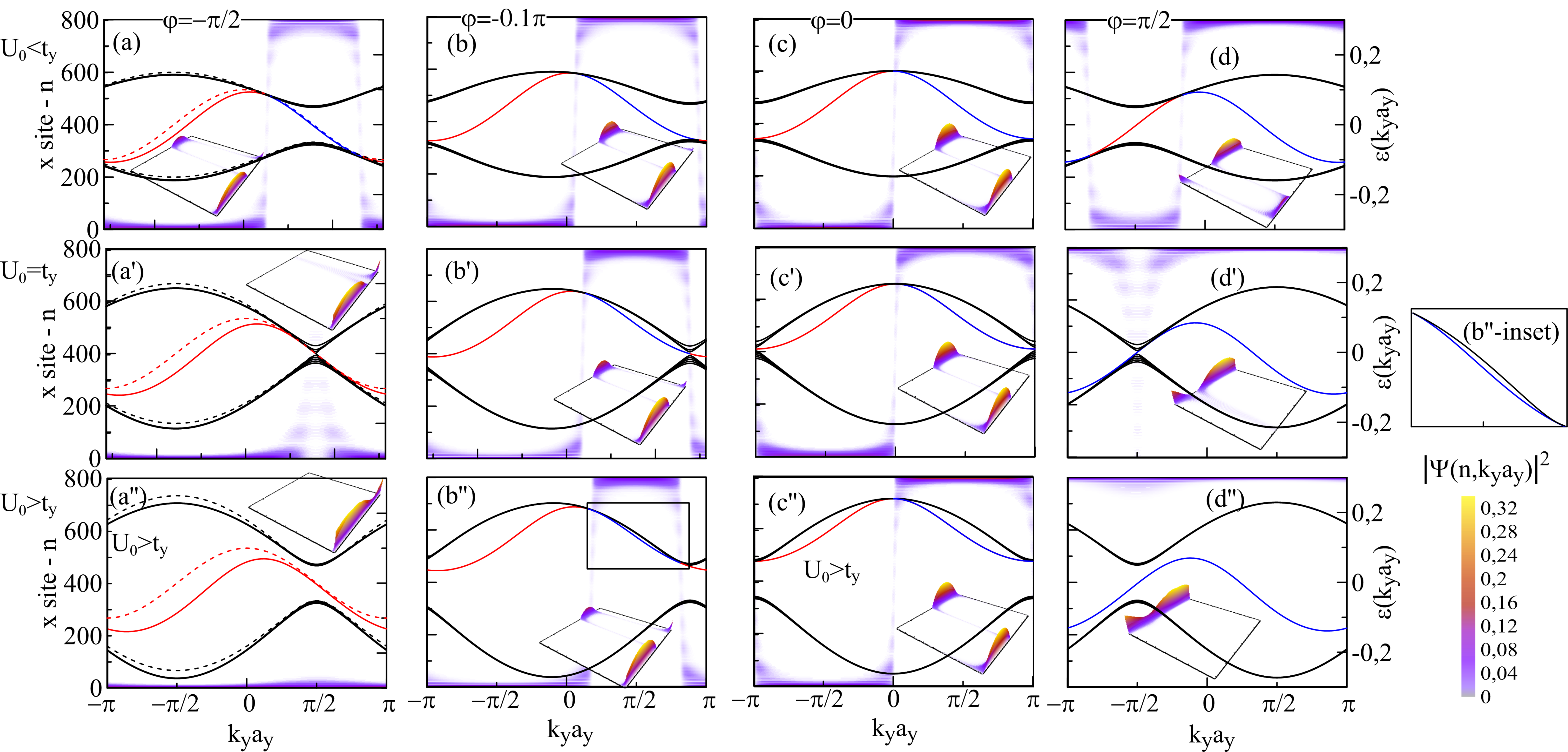}
\caption{\label{fig:nan} The energy spectrum $E(k_y a_y)$ near the band gap for the system (a)-(d) in the topological phase ($t_y>U=0.05t$), (a')-(d') at the phase transition point($t_y=U=0.1t$), (a'')-(d'') in the non-topological  phase ($t_y<U=0.15t$). The panel columns (a), (b), (c), and (d) refer to the phase shift of the CDW modulation  $\varphi=-\pi/2, \ -0.1\pi,\  0,\ \pi/2$, {\it resp}. The energy of the edge state and the bulk spectral edge is found both numerically (solid line) and analytically (dashed line), while the color map represents $|\Psi(n,k_y)|^2$ for the edge state wave function probability. The position at site $n$ (energy) is marked on the left (right) axis.  We note that edge states can be found both in the topological and non-topological phase. By changing $\mu$, the edge states can be tuned between being chiral and non-chiral independent in both phases, see panels (a) and (b'').}
\label{fig:E_WF_phi_0}
\end{figure*}

{\it Continuum model.}  To obtain the analytical solution, it is convenient to change to the continuum 
description~\cite{PhysRevB.82.045127,PhysRevB.86.085408}. In this case, the resonant magnetic field leading to $\phi=2k_{F}a_x$, corresponds to the filling factor $\nu=1$ \cite{PhysRevLett.111.196401}. In the weak tunneling and weak modulation regime, the spectrum can be linearized around the Fermi points, $\pm k_F$, defined via the chemical potential  as $k_F a_x = \arccos (-\mu/2t) $. In the  regime of interest, the CDW modulation competes with the quantum  Hall effect when the period of the CDW is chosen to be  resonant, {\it i.e.}, $k_{w}=k_F$.
The electron operators can be expressed in terms of slowly varying right $[R(x)]$ and left $[L(x)]$ movers as $\Psi(x)=R(x)e^{ik_Fx}+L(x)e^{-ik_Fx}$. The corresponding Hamiltonian density $\mathcal  H$, defined via $H =\int dx\ \Psi^\dagger (x) \mathcal H \Psi (x)$, can be written in terms of  Pauli matrices $\boldsymbol{\sigma}$ acting on the left-right mover subspace $\Psi=(R, L)$ as
\begin{align}
&\mathcal  H= v_F \hat p\sigma_z-\left[U_0\cos{(\varphi)}+t_y\cos{(k_ya_y)}\right]\sigma_x \nonumber \\
    &\hspace{45pt} +\left[U_0\sin{(\varphi)}+t_y\sin{(k_ya_y)}\right]\sigma_y,
\end{align}
where $\hat p  = -i \hbar\partial_x$ is the momentum operator and the Fermi velocity $\upsilon_F$ is given by $\hbar \upsilon_F =2 t a_x\sin(k_Fa_x)$. The bulk energy spectrum is given by
\begin{equation}
E_{\pm}^2=(\hbar v_Fk)^2+U_0^2+t_y^2+2t_yU_0\cos(\varphi-k_ya_y)
\end{equation}
and depends on both $k$ and $k_y$ momenta. Here, $E_+$ ($E_-$) corresponds to the part of the spectrum above (below) $\mu$.
The size of the bulk gap $2\Delta_g={\rm min}_{k}(E_+-E_-)$ for  given $k_y$ becomes
\begin{equation}
    \Delta_g(k_y)\equiv D=\sqrt{U_0^2+t_y^2+2U_0t_y\cos{(\varphi-k_ya_y)}}.
\end{equation}
We note that the system is gapless if $t_y=U_0$ and $\varphi =k_ya_y+\pi$ but fully gapped otherwise, see Fig. 2.  This closing and reopening of the gap  hints to a topological phase transition \cite{alicea2012new}. For a strip of  width $W$, one can thus expect the presence of edge states at the boundaries with energies lying inside the bulk gap. In order to explore the possibility of such edge states we consider a semi-infinite strip ($x\geq 0$) and exploit the method developed in Ref.~\onlinecite{PhysRevLett.109.236801}. Furthermore we assume that $W$ is much larger than the localization length $\xi$ of the edge state \cite{PhysRevB.87.024515}. Thus, we impose vanishing boundary conditions at the end of the strip $\psi_{k_y}(0)=0$, which further imposes the constraint $R(0)=-L(0)$. The energy spectrum of the edge state is then found to be 
\begin{equation}
    \varepsilon (\varphi, k_y)=U_0\cos(\varphi)+t_y\cos(k_ya_y),
\end{equation}
under the condition that $U_0\sin(\varphi)+t_y\sin(k_ya_y)<0$.
The corresponding wavefunction of the left edge state at  energy $\epsilon (\varphi, k_y)$ is given by $\psi_{k_y}(x)\sim\sin(k_Fx)e^{-x/\xi}$
with the localization length
\begin{equation}
    \xi=-\hbar v_F/|U_0\sin(\varphi)+t_y\sin(k_ya_y)|.
	\label{eq:LocalizationLength}
\end{equation}
These edge states propagate along the boundaries in $y$ direction. They can be considered as 1D extension of fractional fermions of the Jackiw-Rebbi type \cite{PhysRevLett.108.136803,PhysRevD.13.3398, PhysRevLett.42.1698, PhysRevB.25.6447,PhysRevLett.109.236801,PhysRevLett.110.126402}.

{\it Topological transition between CDW and QHE phase.}
There are two important phases the system can be tuned into: The non-topological phase, dominated by the CDW modulation, and the topological QHE phase at filling factor  $\nu=1$ (higher filling factors are discussed in the Appendix.~B), dominated by the magnetic field. We study now the transition between these two phases both analytically and by diagonalizing numerically the tight-binding Hamiltonians, see Eqs. (\ref{eq:H_2D}) and (\ref{eq:H_1D_ky}).
In the calculations we fix the  parameters as follows: $t_y/t=0.1$ and $k_F=k_{w}=\pi/4 a_x$. The topological  transition is induced by changing the amplitude of the CDW $U_0$ with respect to the tunneling amplitude $t_y$ between wires. 
We are interested in the bulk band represented by the edge of the gap $\Delta_g(k_y)$ and in the edge state  wave function probability 
$|\psi(n,k_y)|^2$ and its dispersion $\epsilon(k_ya_y)$.

In the topological phase, $t_y>U_0$, the edge state spectrum merges with the bulk gap at two points $\bar{k}_{\pm}$ (one from the electron band and one from the hole band) determined by the condition $\epsilon(\varphi, \bar{k}_\pm)~=~\pm~\Delta_g(\bar{k}_\pm)$,  leading to $\sin (\bar{k}_\pm a_y) =  - (U_0/t_y) \sin \varphi$.
In other words, for any given value of $\varphi$, the edge state exists only for the range of momenta $k_y\in(\bar{k}_{-}, \bar{k}_{+})$, see Fig. 2(a)-(d). 
Here, we can further distinguish between two regimes.  If $\varphi\in (0,\pi)$ [$\varphi\in (-\pi, 0)$] corresponding to the chiral (piecewise chiral) regime, the sign of the Fermi velocity is independent of (depends on) $\mu$, as illustrated  in Fig. 2(a) by the dispersion of the right (chiral) and left (piecewise chiral) edge state. In the piecewise chiral regime, there is a range of $\mu$, for which the edge states are non-chiral, {\it i.e.}, there are two counterpropagating edge modes at a given boundary in contrast to the single edge mode in the chiral regime, where the velocities are opposite at  opposite boundaries. In the topological phase, a non-chiral behaviour is observed for $\mu$ inside the bulk gap.
One can also notice the asymmetry in the localization length between the right and  left edge states. For example, if $\varphi=-\pi/2$, see Fig. 2(a) [$\varphi=\pi/2$, see Fig. 2(d)], the left (right) edge state is more strongly localized than the opposite one which is consistent with  Eq. (\ref{eq:LocalizationLength}) and the 2D finite size calculations [see Fig. 4 (b)] even in the presence of disorder [see Figs. 4 (b') and (b'')]. The larger the gap for given $k_y$ the more localized the edge state is.

In the non-topological phase, $t_y<U_0$,  the edge states exist only for particular values of the CDW phase shift $\varphi$, see Figs. 2(b'') and (d''). Generally, there are three possible scenarios, see Fig. \ref{fig:phase_diagram}. If $t_y< - U_0 \sin \varphi $,
the edge state exists inside the bulk gap without touching the bulk spectrum, see Figs. 2(a'') and (d'').  These edge states are non-chiral and disorder e.g. due to random impurities can result in backscattering inside the same channel, reducing the conductance. If  $t_y< U_0 \sin \varphi $, the system is in the trivial phase without edge states.
In the regime $U_0>t_y>- U_0 \sin \varphi $ ($U_0>t_y>U_0 \sin \varphi $) there are again two wavevectors $k_{\pm}$ at which edge states merge with the bulk electron (hole) spectrum.  As a result, there is a range of chemical potentials (corresponding to the Fermi wavevectors between $k_{-}$ and $k_+$) for which edge states are chiral even in the non-topological phase, see Fig. 2(b''). However, these values are not in the bulk gap, so the edge states coexist with the bulk modes. The previous analysis was relying on the fact that $k_y$ is a good quantum number in the absence of disorder.  Similar to Weyl semimetals \cite{PhysRevB.83.205101, PhysRevB.84.075129, PhysRevLett.107.127205, PhysRevLett.107.186806, PhysRevB.86.115133}, one can expect to detect \cite{PhysRevLett.114.136801} such chiral edge states in a gapless bulk by searching for an enhanced response at the boundaries.

\begin{figure}[ht!]
\centering
\includegraphics[width=8.6cm]{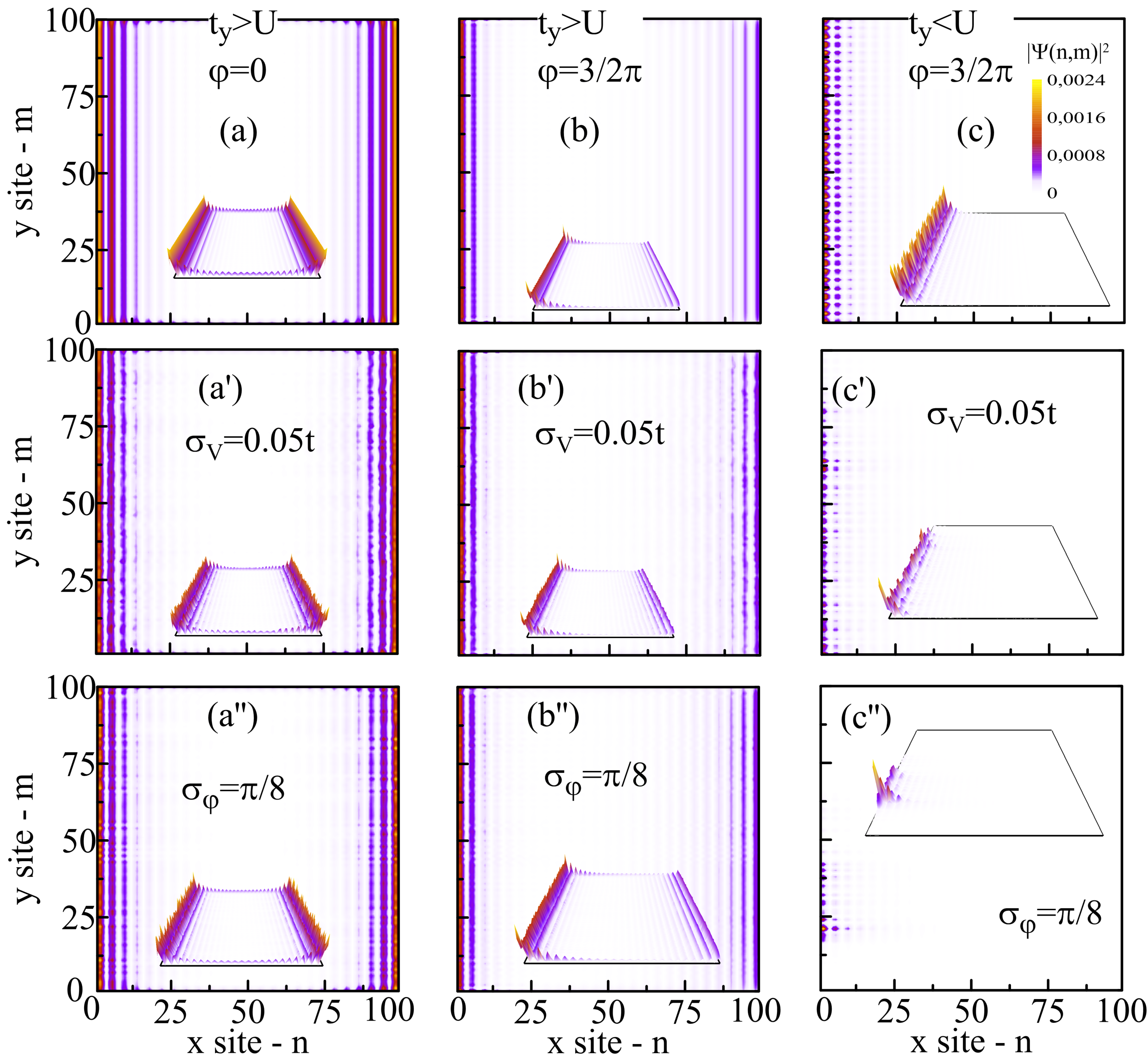}
\caption{\label{fig:2D_CDW_DIS} The wave functions $|\Psi(n,m,E\approx 0)|^2$  of edge states for a 2D finite size lattice inside the bulk gap (near $E=0$) in the topological phase, $t_y>U_0=0.05t$ for (a) $\varphi=0$, (b) $\varphi=3/2\pi$, and (c) in the non-topological phase, $t_y<U_0=0.15t$ and $\varphi=3/2\pi$. The lower panels $(a')$-$(c')$  and $(a'')$-$(c'')$ correspond to the case of onsite disorder ($\sigma_{V}=0.05t$) and disorder in the CDW phase ($\sigma_{\varphi}=\pi/8$). In the topological phase, the presence of edge states is not affected by weak disorder (with effective amplitude not exceeding the size of the gap), while edge states in the non-topological phase are affected by weak disorder leading to  Anderson localization (around some random edge site).  However, even in the latter case edge states can survive some small amount of disorder of both types when  $\sigma_{V}\apprle 0.1t$ and $\sigma_{V}\apprle\pi/16$.}
\label{fig:E_WF_phi_0}
\end{figure}

{\it Disorder effects.}
In realistic systems one cannot avoid disorder. In our 2D finite size lattice model, we study effects of disorder by introducing (i) a random on-site potential $\sum_{n,m}V_{n,m}c_{n,m}^{\dag}c_{n,m}$ and (ii) a random phase $\tilde{\varphi}_m$ for the CDW modulation in each wire, {\it i.e.},
$2U_0\sum_{n,m}\cos(2k_{w}na_x+\varphi+\tilde{\varphi}_m)c^{\dag}_{n,m}c_{n,m}$. Here, $V_{n,m}$ ($\tilde{\varphi}_m$) is taken according to a Gaussian distribution with zero mean and standard deviation $\sigma_V$ ($\sigma_{\varphi}$).
By analyzing the edge state wave functions $|\Psi(n,m,\varphi)|^2$ in different phases (see  Fig. 4), we see that, for the given variance $\sigma_V\lesssim 0.1t$, the onsite disorder does not destroy the edge states in both topological [also the asymmetry in localization lengths is preserved, see Fig. 4 (b')] and non-topological phases, see Figs. 4 (a')-(c') \footnote{One can set the system parameters and disorder strength in such a way that only QHE edge states will survive.}.
Interestingly, chiral QHE edge states survive any amount of disorder in the phase $\varphi$, while in the non-topological phase, the edge states survive only up to a certain small amount of disorder $\sigma_{\varphi}$ with stronger disorder leading to Anderson localization around a random location along the edge.

\begin{figure}[H]
\centering
\vspace{2pt}
\includegraphics[width=8.6cm]{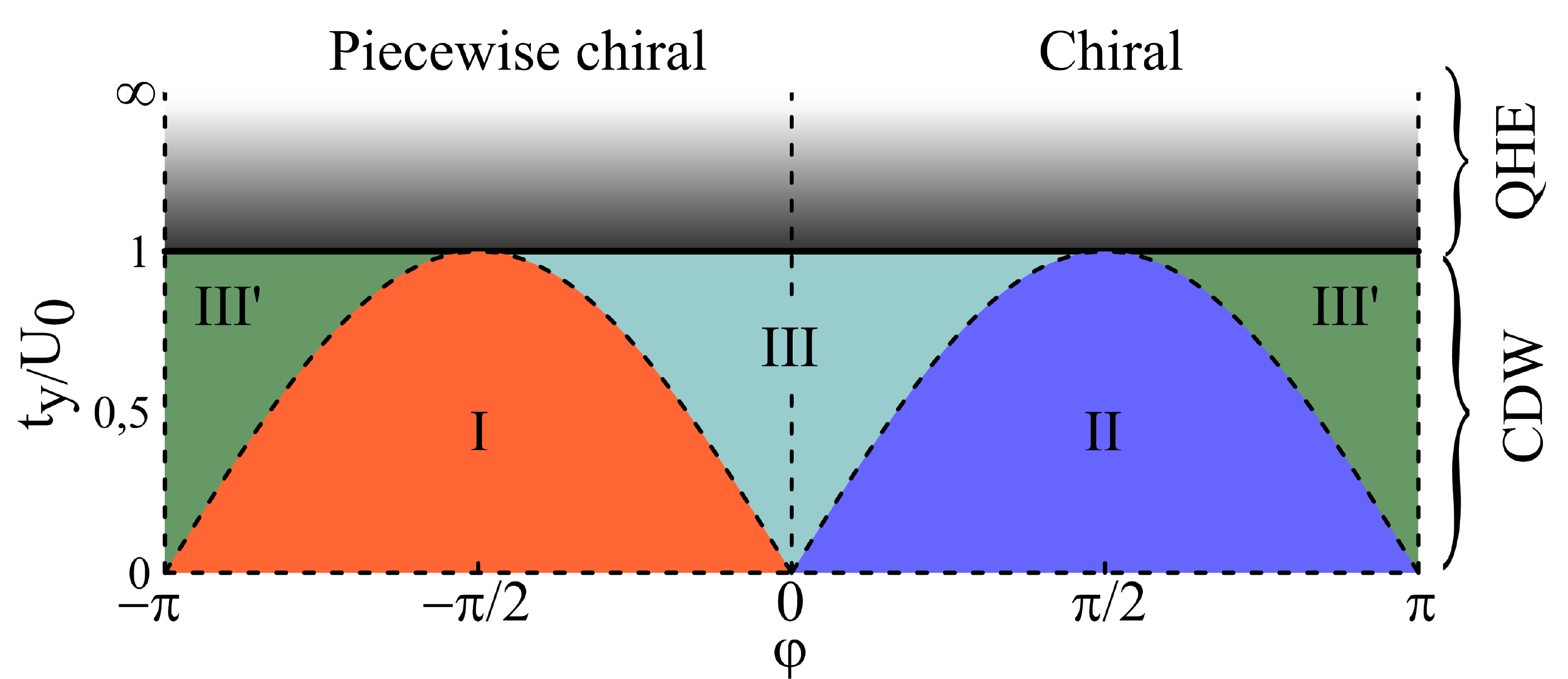}
\caption{Phase diagram for filling factor $\nu=1$. If $t_y/U_0>1$ ($t_y/U_0<1$), the system is in the topological (non-topological) phase, indicated as QHE (CDW) phase.  The phase transition between phases occurs at $t_y/U_0=1$, where the gap closes. The non-topological phase is subdivided into following subphases: (I) (orange area) edge states are totally inside the bulk gap,  (II) (blue area) no edge states, and edge states spread between two wavevectors $k_-$ and $k_+$ belonging either both to the electron (III) (light green area) or both to the hole (III') (dark green areas) band. 
At the phase boundary between the subphases (dashed lines), we have $k_-=k_+$. 
Generally, depending on the phase $\varphi$ the edge states can be either chiral for all values of $\mu$ or be piecewise chiral, such that by shifting $\mu$, opposite chiralities can be observed. }
\label{fig:phase_diagram}
\end{figure}
{\it Summary.} 
We have studied a system of weakly coupled and CDW modulated wires in a perpendicular magnetic field. The  system supports edge states in both the non-topological (CDW dominated) and topological (QHE dominated) phase. Interestingly, both phases host chiral and non-chiral edge states depending on the chemical potential position.
Numerical calculations showed that, in general, the edge states in the non-topological phase are more affected by the disorder than in the topological phase, however, the former can still survive a finite amount of disorder. 
We propose that our predictions can be tested in semiconducting nanowires with CDW modulations, but also in optical lattices~\cite{lewenstein2007ultracold,PhysRevLett.111.185301,aidelsburger2015measuring} or  photonic crystals~\cite{PhysRevLett.109.106402}. 
Finally, it would be interesting to see if our results can be extended to other models which include TI phases  with (piecewise) helical edge states.

\begin{acknowledgments}
We acknowledge support from the Swiss NSF, NCCR QSIT, and SCIEX.
\end{acknowledgments}
 
\nocite{*}

\bibliography{QH_CDW_PRL}

\providecommand{\noopsort}[1]{}\providecommand{\singleletter}[1]{#1}%
\begin{thebibliography}{48}%
\makeatletter
\providecommand \@ifxundefined [1]{%
 \@ifx{#1\undefined}
}%
\providecommand \@ifnum [1]{%
 \ifnum #1\expandafter \@firstoftwo
 \else \expandafter \@secondoftwo
 \fi
}%
\providecommand \@ifx [1]{%
 \ifx #1\expandafter \@firstoftwo
 \else \expandafter \@secondoftwo
 \fi
}%
\providecommand \natexlab [1]{#1}%
\providecommand \enquote  [1]{``#1''}%
\providecommand \bibnamefont  [1]{#1}%
\providecommand \bibfnamefont [1]{#1}%
\providecommand \citenamefont [1]{#1}%
\providecommand \href@noop [0]{\@secondoftwo}%
\providecommand \href [0]{\begingroup \@sanitize@url \@href}%
\providecommand \@href[1]{\@@startlink{#1}\@@href}%
\providecommand \@@href[1]{\endgroup#1\@@endlink}%
\providecommand \@sanitize@url [0]{\catcode `\\12\catcode `\$12\catcode
  `\&12\catcode `\#12\catcode `\^12\catcode `\_12\catcode `\%12\relax}%
\providecommand \@@startlink[1]{}%
\providecommand \@@endlink[0]{}%
\providecommand \url  [0]{\begingroup\@sanitize@url \@url }%
\providecommand \@url [1]{\endgroup\@href {#1}{\urlprefix }}%
\providecommand \urlprefix  [0]{URL }%
\providecommand \Eprint [0]{\href }%
\providecommand \doibase [0]{http://dx.doi.org/}%
\providecommand \selectlanguage [0]{\@gobble}%
\providecommand \bibinfo  [0]{\@secondoftwo}%
\providecommand \bibfield  [0]{\@secondoftwo}%
\providecommand \translation [1]{[#1]}%
\providecommand \BibitemOpen [0]{}%
\providecommand \bibitemStop [0]{}%
\providecommand \bibitemNoStop [0]{.\EOS\space}%
\providecommand \EOS [0]{\spacefactor3000\relax}%
\providecommand \BibitemShut  [1]{\csname bibitem#1\endcsname}%
\let\auto@bib@innerbib\@empty
\bibitem [{\citenamefont {Klitzing}\ \emph {et~al.}(1980)\citenamefont
  {Klitzing}, \citenamefont {Dorda},\ and\ \citenamefont
  {Pepper}}]{PhysRevLett.45.494}%
  \BibitemOpen
  \bibfield  {author} {\bibinfo {author} {\bibfnamefont {K.~v.}\ \bibnamefont
  {Klitzing}}, \bibinfo {author} {\bibfnamefont {G.}~\bibnamefont {Dorda}}, \
  and\ \bibinfo {author} {\bibfnamefont {M.}~\bibnamefont {Pepper}},\ }\href
  {\doibase 10.1103/PhysRevLett.45.494} {\bibfield  {journal} {\bibinfo
  {journal} {Phys. Rev. Lett.}\ }\textbf {\bibinfo {volume} {45}},\ \bibinfo
  {pages} {494} (\bibinfo {year} {1980})}\BibitemShut {NoStop}%
\bibitem [{\citenamefont {Tsui}\ \emph {et~al.}(1982)\citenamefont {Tsui},
  \citenamefont {Stormer},\ and\ \citenamefont
  {Gossard}}]{PhysRevLett.48.1559}%
  \BibitemOpen
  \bibfield  {author} {\bibinfo {author} {\bibfnamefont {D.~C.}\ \bibnamefont
  {Tsui}}, \bibinfo {author} {\bibfnamefont {H.~L.}\ \bibnamefont {Stormer}}, \
  and\ \bibinfo {author} {\bibfnamefont {A.~C.}\ \bibnamefont {Gossard}},\
  }\href {\doibase 10.1103/PhysRevLett.48.1559} {\bibfield  {journal} {\bibinfo
   {journal} {Phys. Rev. Lett.}\ }\textbf {\bibinfo {volume} {48}},\ \bibinfo
  {pages} {1559} (\bibinfo {year} {1982})}\BibitemShut {NoStop}%
\bibitem [{\citenamefont {Thouless}\ \emph {et~al.}(1982)\citenamefont
  {Thouless}, \citenamefont {Kohmoto}, \citenamefont {Nightingale},\ and\
  \citenamefont {den Nijs}}]{PhysRevLett.49.405}%
  \BibitemOpen
  \bibfield  {author} {\bibinfo {author} {\bibfnamefont {D.~J.}\ \bibnamefont
  {Thouless}}, \bibinfo {author} {\bibfnamefont {M.}~\bibnamefont {Kohmoto}},
  \bibinfo {author} {\bibfnamefont {M.~P.}\ \bibnamefont {Nightingale}}, \ and\
  \bibinfo {author} {\bibfnamefont {M.}~\bibnamefont {den Nijs}},\ }\href
  {\doibase 10.1103/PhysRevLett.49.405} {\bibfield  {journal} {\bibinfo
  {journal} {Phys. Rev. Lett.}\ }\textbf {\bibinfo {volume} {49}},\ \bibinfo
  {pages} {405} (\bibinfo {year} {1982})}\BibitemShut {NoStop}%
\bibitem [{\citenamefont {Hasan}\ and\ \citenamefont
  {Kane}(2010)}]{RevModPhys.82.3045}%
  \BibitemOpen
  \bibfield  {author} {\bibinfo {author} {\bibfnamefont {M.~Z.}\ \bibnamefont
  {Hasan}}\ and\ \bibinfo {author} {\bibfnamefont {C.~L.}\ \bibnamefont
  {Kane}},\ }\href {\doibase 10.1103/RevModPhys.82.3045} {\bibfield  {journal}
  {\bibinfo  {journal} {Rev. Mod. Phys.}\ }\textbf {\bibinfo {volume} {82}},\
  \bibinfo {pages} {3045} (\bibinfo {year} {2010})}\BibitemShut {NoStop}%
\bibitem [{\citenamefont {Hart}\ \emph {et~al.}(2014)\citenamefont {Hart},
  \citenamefont {Ren}, \citenamefont {Wagner}, \citenamefont {Leubner},
  \citenamefont {M{\"u}hlbauer}, \citenamefont {Br{\"u}ne}, \citenamefont
  {Buhmann}, \citenamefont {Molenkamp},\ and\ \citenamefont
  {Yacoby}}]{hart2014induced}%
  \BibitemOpen
  \bibfield  {author} {\bibinfo {author} {\bibfnamefont {S.}~\bibnamefont
  {Hart}}, \bibinfo {author} {\bibfnamefont {H.}~\bibnamefont {Ren}}, \bibinfo
  {author} {\bibfnamefont {T.}~\bibnamefont {Wagner}}, \bibinfo {author}
  {\bibfnamefont {P.}~\bibnamefont {Leubner}}, \bibinfo {author} {\bibfnamefont
  {M.}~\bibnamefont {M{\"u}hlbauer}}, \bibinfo {author} {\bibfnamefont
  {C.}~\bibnamefont {Br{\"u}ne}}, \bibinfo {author} {\bibfnamefont
  {H.}~\bibnamefont {Buhmann}}, \bibinfo {author} {\bibfnamefont {L.~W.}\
  \bibnamefont {Molenkamp}}, \ and\ \bibinfo {author} {\bibfnamefont
  {A.}~\bibnamefont {Yacoby}},\ }\href@noop {} {\bibfield  {journal} {\bibinfo
  {journal} {Nature Physics}\ } (\bibinfo {year} {2014})}\BibitemShut {NoStop}%
\bibitem [{\citenamefont {Pribiag}\ \emph {et~al.}(2015)\citenamefont
  {Pribiag}, \citenamefont {Beukman}, \citenamefont {Qu}, \citenamefont
  {Cassidy}, \citenamefont {Charpentier}, \citenamefont {Wegscheider},\ and\
  \citenamefont {Kouwenhoven}}]{pribiag2015edge}%
  \BibitemOpen
  \bibfield  {author} {\bibinfo {author} {\bibfnamefont {V.~S.}\ \bibnamefont
  {Pribiag}}, \bibinfo {author} {\bibfnamefont {A.~J.}\ \bibnamefont
  {Beukman}}, \bibinfo {author} {\bibfnamefont {F.}~\bibnamefont {Qu}},
  \bibinfo {author} {\bibfnamefont {M.~C.}\ \bibnamefont {Cassidy}}, \bibinfo
  {author} {\bibfnamefont {C.}~\bibnamefont {Charpentier}}, \bibinfo {author}
  {\bibfnamefont {W.}~\bibnamefont {Wegscheider}}, \ and\ \bibinfo {author}
  {\bibfnamefont {L.~P.}\ \bibnamefont {Kouwenhoven}},\ }\href@noop {}
  {\bibfield  {journal} {\bibinfo  {journal} {Nature nanotechnology}\ }
  (\bibinfo {year} {2015})}\BibitemShut {NoStop}%
\bibitem [{\citenamefont {Spanton}\ \emph {et~al.}(2014)\citenamefont
  {Spanton}, \citenamefont {Nowack}, \citenamefont {Du}, \citenamefont
  {Sullivan}, \citenamefont {Du},\ and\ \citenamefont
  {Moler}}]{PhysRevLett.113.026804}%
  \BibitemOpen
  \bibfield  {author} {\bibinfo {author} {\bibfnamefont {E.~M.}\ \bibnamefont
  {Spanton}}, \bibinfo {author} {\bibfnamefont {K.~C.}\ \bibnamefont {Nowack}},
  \bibinfo {author} {\bibfnamefont {L.}~\bibnamefont {Du}}, \bibinfo {author}
  {\bibfnamefont {G.}~\bibnamefont {Sullivan}}, \bibinfo {author}
  {\bibfnamefont {R.-R.}\ \bibnamefont {Du}}, \ and\ \bibinfo {author}
  {\bibfnamefont {K.~A.}\ \bibnamefont {Moler}},\ }\href {\doibase
  10.1103/PhysRevLett.113.026804} {\bibfield  {journal} {\bibinfo  {journal}
  {Phys. Rev. Lett.}\ }\textbf {\bibinfo {volume} {113}},\ \bibinfo {pages}
  {026804} (\bibinfo {year} {2014})}\BibitemShut {NoStop}%
\bibitem [{\citenamefont {Wang}\ \emph {et~al.}(2015)\citenamefont {Wang},
  \citenamefont {Kirtley}, \citenamefont {Katmis}, \citenamefont
  {Jarillo-Herrero}, \citenamefont {Moodera},\ and\ \citenamefont
  {Moler}}]{Wang28082015}%
  \BibitemOpen
  \bibfield  {author} {\bibinfo {author} {\bibfnamefont {Y.~H.}\ \bibnamefont
  {Wang}}, \bibinfo {author} {\bibfnamefont {J.~R.}\ \bibnamefont {Kirtley}},
  \bibinfo {author} {\bibfnamefont {F.}~\bibnamefont {Katmis}}, \bibinfo
  {author} {\bibfnamefont {P.}~\bibnamefont {Jarillo-Herrero}}, \bibinfo
  {author} {\bibfnamefont {J.~S.}\ \bibnamefont {Moodera}}, \ and\ \bibinfo
  {author} {\bibfnamefont {K.~A.}\ \bibnamefont {Moler}},\ }\href {\doibase
  10.1126/science.aaa0508} {\bibfield  {journal} {\bibinfo  {journal}
  {Science}\ }\textbf {\bibinfo {volume} {349}},\ \bibinfo {pages} {948}
  (\bibinfo {year} {2015})}\BibitemShut {NoStop}%
\bibitem [{\citenamefont {Tamm}(1932)}]{ITamm}%
  \BibitemOpen
  \bibfield  {author} {\bibinfo {author} {\bibfnamefont {I.}~\bibnamefont
  {Tamm}},\ }\href@noop {} {\bibfield  {journal} {\bibinfo  {journal} {Z.
  Sowjetunion 1}\ } (\bibinfo {year} {1932})}\BibitemShut {NoStop}%
\bibitem [{\citenamefont {Shockley}(1939)}]{PhysRev.56.317}%
  \BibitemOpen
  \bibfield  {author} {\bibinfo {author} {\bibfnamefont {W.}~\bibnamefont
  {Shockley}},\ }\href {\doibase 10.1103/PhysRev.56.317} {\bibfield  {journal}
  {\bibinfo  {journal} {Phys. Rev.}\ }\textbf {\bibinfo {volume} {56}},\
  \bibinfo {pages} {317} (\bibinfo {year} {1939})}\BibitemShut {NoStop}%
\bibitem [{\citenamefont {Gangadharaiah}\ \emph {et~al.}(2012)\citenamefont
  {Gangadharaiah}, \citenamefont {Trifunovic},\ and\ \citenamefont
  {Loss}}]{PhysRevLett.108.136803}%
  \BibitemOpen
  \bibfield  {author} {\bibinfo {author} {\bibfnamefont {S.}~\bibnamefont
  {Gangadharaiah}}, \bibinfo {author} {\bibfnamefont {L.}~\bibnamefont
  {Trifunovic}}, \ and\ \bibinfo {author} {\bibfnamefont {D.}~\bibnamefont
  {Loss}},\ }\href {\doibase 10.1103/PhysRevLett.108.136803} {\bibfield
  {journal} {\bibinfo  {journal} {Phys. Rev. Lett.}\ }\textbf {\bibinfo
  {volume} {108}},\ \bibinfo {pages} {136803} (\bibinfo {year}
  {2012})}\BibitemShut {NoStop}%
\bibitem [{\citenamefont {Nichele}\ \emph {et~al.}(2015)\citenamefont
  {Nichele}, \citenamefont {Suominen}, \citenamefont {Kjaergaard},
  \citenamefont {Marcus}, \citenamefont {Sajadi}, \citenamefont {Folk},
  \citenamefont {Qu}, \citenamefont {Beukman}, \citenamefont {Vries},
  \citenamefont {Veen}, \citenamefont {Nadj-Perge}, \citenamefont
  {Kouwenhoven}, \citenamefont {Nguyen}, \citenamefont {Kiselev}, \citenamefont
  {Yi}, \citenamefont {Sokolich}, \citenamefont {Manfra}, \citenamefont
  {Spanton},\ and\ \citenamefont {Moler}}]{Nichele2015}%
  \BibitemOpen
  \bibfield  {author} {\bibinfo {author} {\bibfnamefont {F.}~\bibnamefont
  {Nichele}}, \bibinfo {author} {\bibfnamefont {H.~J.}\ \bibnamefont
  {Suominen}}, \bibinfo {author} {\bibfnamefont {M.}~\bibnamefont
  {Kjaergaard}}, \bibinfo {author} {\bibfnamefont {C.~M.}\ \bibnamefont
  {Marcus}}, \bibinfo {author} {\bibfnamefont {E.}~\bibnamefont {Sajadi}},
  \bibinfo {author} {\bibfnamefont {J.~A.}\ \bibnamefont {Folk}}, \bibinfo
  {author} {\bibfnamefont {F.}~\bibnamefont {Qu}}, \bibinfo {author}
  {\bibfnamefont {A.~J.}\ \bibnamefont {Beukman}}, \bibinfo {author}
  {\bibfnamefont {F.~K.~d.}\ \bibnamefont {Vries}}, \bibinfo {author}
  {\bibfnamefont {J.~v.}\ \bibnamefont {Veen}}, \bibinfo {author}
  {\bibfnamefont {S.}~\bibnamefont {Nadj-Perge}}, \bibinfo {author}
  {\bibfnamefont {L.~P.}\ \bibnamefont {Kouwenhoven}}, \bibinfo {author}
  {\bibfnamefont {B.-M.}\ \bibnamefont {Nguyen}}, \bibinfo {author}
  {\bibfnamefont {A.~A.}\ \bibnamefont {Kiselev}}, \bibinfo {author}
  {\bibfnamefont {W.}~\bibnamefont {Yi}}, \bibinfo {author} {\bibfnamefont
  {M.}~\bibnamefont {Sokolich}}, \bibinfo {author} {\bibfnamefont {M.~J.}\
  \bibnamefont {Manfra}}, \bibinfo {author} {\bibfnamefont {E.~M.}\
  \bibnamefont {Spanton}}, \ and\ \bibinfo {author} {\bibfnamefont {K.~A.}\
  \bibnamefont {Moler}},\ }\href {shttp://arxiv.org/abs/1511.01728} {\bibfield
  {journal} {\bibinfo  {journal} {arXiv:1511.01728}\ } (\bibinfo {year}
  {2015})}\BibitemShut {NoStop}%
\bibitem [{\citenamefont {Giamarchi}(2003)}]{giamarchi2003quantum}%
  \BibitemOpen
  \bibfield  {author} {\bibinfo {author} {\bibfnamefont {T.}~\bibnamefont
  {Giamarchi}},\ }\href {https://books.google.ch/books?id=GVeuKZLGMZ0C} {\emph
  {\bibinfo {title} {Quantum Physics in One Dimension}}},\ International Series
  of Monographs on Physics\ (\bibinfo  {publisher} {Clarendon Press},\ \bibinfo
  {year} {2003})\BibitemShut {NoStop}%
\bibitem [{\citenamefont {Klinovaja}\ and\ \citenamefont
  {Loss}(2014)}]{FQHE_CDW_Jelena}%
  \BibitemOpen
  \bibfield  {author} {\bibinfo {author} {\bibfnamefont {J.}~\bibnamefont
  {Klinovaja}}\ and\ \bibinfo {author} {\bibfnamefont {D.}~\bibnamefont
  {Loss}},\ }\href {\doibase 10.1140/epjb/e2014-50395-6} {\bibfield  {journal}
  {\bibinfo  {journal} {The European Physical Journal B}\ }\textbf {\bibinfo
  {volume} {87}},\ \bibinfo {eid} {171} (\bibinfo {year} {2014}),\
  10.1140/epjb/e2014-50395-6}\BibitemShut {NoStop}%
\bibitem [{\citenamefont {{Bedell}}(1994)}]{1994lanm.proc...15B}%
  \BibitemOpen
  \bibfield  {author} {\bibinfo {author} {\bibfnamefont {K.~S.}\ \bibnamefont
  {{Bedell}}},\ }in\ \href@noop {} {\emph {\bibinfo {booktitle} {Proceedings
  held in Los Alamos, NM, 15-18 Dec. 1993}}}\ (\bibinfo {year} {1994})\ pp.\
  \bibinfo {pages} {15--18}\BibitemShut {NoStop}%
\bibitem [{\citenamefont {Algra}\ \emph {et~al.}(2008)\citenamefont {Algra},
  \citenamefont {Verheijen}, \citenamefont {Borgstr{\"o}m}, \citenamefont
  {Feiner}, \citenamefont {Immink}, \citenamefont {van Enckevort},
  \citenamefont {Vlieg},\ and\ \citenamefont {Bakkers}}]{algra2008twinning}%
  \BibitemOpen
  \bibfield  {author} {\bibinfo {author} {\bibfnamefont {R.~E.}\ \bibnamefont
  {Algra}}, \bibinfo {author} {\bibfnamefont {M.~A.}\ \bibnamefont
  {Verheijen}}, \bibinfo {author} {\bibfnamefont {M.~T.}\ \bibnamefont
  {Borgstr{\"o}m}}, \bibinfo {author} {\bibfnamefont {L.-F.}\ \bibnamefont
  {Feiner}}, \bibinfo {author} {\bibfnamefont {G.}~\bibnamefont {Immink}},
  \bibinfo {author} {\bibfnamefont {W.~J.}\ \bibnamefont {van Enckevort}},
  \bibinfo {author} {\bibfnamefont {E.}~\bibnamefont {Vlieg}}, \ and\ \bibinfo
  {author} {\bibfnamefont {E.~P.}\ \bibnamefont {Bakkers}},\ }\href@noop {}
  {\bibfield  {journal} {\bibinfo  {journal} {Nature}\ }\textbf {\bibinfo
  {volume} {456}},\ \bibinfo {pages} {369} (\bibinfo {year}
  {2008})}\BibitemShut {NoStop}%
\bibitem [{\citenamefont {Deutschmann}\ \emph {et~al.}(2001)\citenamefont
  {Deutschmann}, \citenamefont {Wegscheider}, \citenamefont {Rother},
  \citenamefont {Bichler}, \citenamefont {Abstreiter}, \citenamefont
  {Albrecht},\ and\ \citenamefont {Smet}}]{PhysRevLett.86.1857}%
  \BibitemOpen
  \bibfield  {author} {\bibinfo {author} {\bibfnamefont {R.~A.}\ \bibnamefont
  {Deutschmann}}, \bibinfo {author} {\bibfnamefont {W.}~\bibnamefont
  {Wegscheider}}, \bibinfo {author} {\bibfnamefont {M.}~\bibnamefont {Rother}},
  \bibinfo {author} {\bibfnamefont {M.}~\bibnamefont {Bichler}}, \bibinfo
  {author} {\bibfnamefont {G.}~\bibnamefont {Abstreiter}}, \bibinfo {author}
  {\bibfnamefont {C.}~\bibnamefont {Albrecht}}, \ and\ \bibinfo {author}
  {\bibfnamefont {J.~H.}\ \bibnamefont {Smet}},\ }\href {\doibase
  10.1103/PhysRevLett.86.1857} {\bibfield  {journal} {\bibinfo  {journal}
  {Phys. Rev. Lett.}\ }\textbf {\bibinfo {volume} {86}},\ \bibinfo {pages}
  {1857} (\bibinfo {year} {2001})}\BibitemShut {NoStop}%
\bibitem [{\citenamefont {Lebed}(1986)}]{lebed1986anisotropy}%
  \BibitemOpen
  \bibfield  {author} {\bibinfo {author} {\bibfnamefont {A.}~\bibnamefont
  {Lebed}},\ }\href@noop {} {\bibfield  {journal} {\bibinfo  {journal} {ZhETF
  Pisma Redaktsiiu}\ }\textbf {\bibinfo {volume} {43}},\ \bibinfo {pages} {137}
  (\bibinfo {year} {1986})}\BibitemShut {NoStop}%
\bibitem [{\citenamefont {Yakovenko}(1991)}]{PhysRevB.43.11353}%
  \BibitemOpen
  \bibfield  {author} {\bibinfo {author} {\bibfnamefont {V.~M.}\ \bibnamefont
  {Yakovenko}},\ }\href {\doibase 10.1103/PhysRevB.43.11353} {\bibfield
  {journal} {\bibinfo  {journal} {Phys. Rev. B}\ }\textbf {\bibinfo {volume}
  {43}},\ \bibinfo {pages} {11353} (\bibinfo {year} {1991})}\BibitemShut
  {NoStop}%
\bibitem [{\citenamefont {Klinovaja}\ and\ \citenamefont
  {Loss}(2013{\natexlab{a}})}]{PhysRevLett.111.196401}%
  \BibitemOpen
  \bibfield  {author} {\bibinfo {author} {\bibfnamefont {J.}~\bibnamefont
  {Klinovaja}}\ and\ \bibinfo {author} {\bibfnamefont {D.}~\bibnamefont
  {Loss}},\ }\href {\doibase 10.1103/PhysRevLett.111.196401} {\bibfield
  {journal} {\bibinfo  {journal} {Phys. Rev. Lett.}\ }\textbf {\bibinfo
  {volume} {111}},\ \bibinfo {pages} {196401} (\bibinfo {year}
  {2013}{\natexlab{a}})}\BibitemShut {NoStop}%
\bibitem [{\citenamefont {Teo}\ and\ \citenamefont
  {Kane}(2014)}]{PhysRevB.89.085101}%
  \BibitemOpen
  \bibfield  {author} {\bibinfo {author} {\bibfnamefont {J.~C.~Y.}\
  \bibnamefont {Teo}}\ and\ \bibinfo {author} {\bibfnamefont {C.~L.}\
  \bibnamefont {Kane}},\ }\href {\doibase 10.1103/PhysRevB.89.085101}
  {\bibfield  {journal} {\bibinfo  {journal} {Phys. Rev. B}\ }\textbf {\bibinfo
  {volume} {89}},\ \bibinfo {pages} {085101} (\bibinfo {year}
  {2014})}\BibitemShut {NoStop}%
\bibitem [{\citenamefont {Klinovaja}\ and\ \citenamefont
  {Tserkovnyak}(2014)}]{PhysRevB.90.115426}%
  \BibitemOpen
  \bibfield  {author} {\bibinfo {author} {\bibfnamefont {J.}~\bibnamefont
  {Klinovaja}}\ and\ \bibinfo {author} {\bibfnamefont {Y.}~\bibnamefont
  {Tserkovnyak}},\ }\href {\doibase 10.1103/PhysRevB.90.115426} {\bibfield
  {journal} {\bibinfo  {journal} {Phys. Rev. B}\ }\textbf {\bibinfo {volume}
  {90}},\ \bibinfo {pages} {115426} (\bibinfo {year} {2014})}\BibitemShut
  {NoStop}%
\bibitem [{\citenamefont {Seroussi}\ \emph {et~al.}(2014)\citenamefont
  {Seroussi}, \citenamefont {Berg},\ and\ \citenamefont
  {Oreg}}]{PhysRevB.89.104523}%
  \BibitemOpen
  \bibfield  {author} {\bibinfo {author} {\bibfnamefont {I.}~\bibnamefont
  {Seroussi}}, \bibinfo {author} {\bibfnamefont {E.}~\bibnamefont {Berg}}, \
  and\ \bibinfo {author} {\bibfnamefont {Y.}~\bibnamefont {Oreg}},\ }\href
  {\doibase 10.1103/PhysRevB.89.104523} {\bibfield  {journal} {\bibinfo
  {journal} {Phys. Rev. B}\ }\textbf {\bibinfo {volume} {89}},\ \bibinfo
  {pages} {104523} (\bibinfo {year} {2014})}\BibitemShut {NoStop}%
\bibitem [{\citenamefont {Sagi}\ and\ \citenamefont
  {Oreg}(2014)}]{PhysRevB.90.201102}%
  \BibitemOpen
  \bibfield  {author} {\bibinfo {author} {\bibfnamefont {E.}~\bibnamefont
  {Sagi}}\ and\ \bibinfo {author} {\bibfnamefont {Y.}~\bibnamefont {Oreg}},\
  }\href {\doibase 10.1103/PhysRevB.90.201102} {\bibfield  {journal} {\bibinfo
  {journal} {Phys. Rev. B}\ }\textbf {\bibinfo {volume} {90}},\ \bibinfo
  {pages} {201102} (\bibinfo {year} {2014})}\BibitemShut {NoStop}%
\bibitem [{\citenamefont {Meng}\ and\ \citenamefont
  {Sela}(2014)}]{PhysRevB.90.235425}%
  \BibitemOpen
  \bibfield  {author} {\bibinfo {author} {\bibfnamefont {T.}~\bibnamefont
  {Meng}}\ and\ \bibinfo {author} {\bibfnamefont {E.}~\bibnamefont {Sela}},\
  }\href {\doibase 10.1103/PhysRevB.90.235425} {\bibfield  {journal} {\bibinfo
  {journal} {Phys. Rev. B}\ }\textbf {\bibinfo {volume} {90}},\ \bibinfo
  {pages} {235425} (\bibinfo {year} {2014})}\BibitemShut {NoStop}%
\bibitem [{\citenamefont {Klinovaja}\ \emph {et~al.}(2015)\citenamefont
  {Klinovaja}, \citenamefont {Tserkovnyak},\ and\ \citenamefont
  {Loss}}]{PhysRevB.91.085426}%
  \BibitemOpen
  \bibfield  {author} {\bibinfo {author} {\bibfnamefont {J.}~\bibnamefont
  {Klinovaja}}, \bibinfo {author} {\bibfnamefont {Y.}~\bibnamefont
  {Tserkovnyak}}, \ and\ \bibinfo {author} {\bibfnamefont {D.}~\bibnamefont
  {Loss}},\ }\href {\doibase 10.1103/PhysRevB.91.085426} {\bibfield  {journal}
  {\bibinfo  {journal} {Phys. Rev. B}\ }\textbf {\bibinfo {volume} {91}},\
  \bibinfo {pages} {085426} (\bibinfo {year} {2015})}\BibitemShut {NoStop}%
\bibitem [{\citenamefont {Meng}\ \emph {et~al.}(2015)\citenamefont {Meng},
  \citenamefont {Neupert}, \citenamefont {Greiter},\ and\ \citenamefont
  {Thomale}}]{PhysRevB.91.241106}%
  \BibitemOpen
  \bibfield  {author} {\bibinfo {author} {\bibfnamefont {T.}~\bibnamefont
  {Meng}}, \bibinfo {author} {\bibfnamefont {T.}~\bibnamefont {Neupert}},
  \bibinfo {author} {\bibfnamefont {M.}~\bibnamefont {Greiter}}, \ and\
  \bibinfo {author} {\bibfnamefont {R.}~\bibnamefont {Thomale}},\ }\href
  {\doibase 10.1103/PhysRevB.91.241106} {\bibfield  {journal} {\bibinfo
  {journal} {Phys. Rev. B}\ }\textbf {\bibinfo {volume} {91}},\ \bibinfo
  {pages} {241106} (\bibinfo {year} {2015})}\BibitemShut {NoStop}%
\bibitem [{\citenamefont {Braunecker}\ \emph {et~al.}(2010)\citenamefont
  {Braunecker}, \citenamefont {Japaridze}, \citenamefont {Klinovaja},\ and\
  \citenamefont {Loss}}]{PhysRevB.82.045127}%
  \BibitemOpen
  \bibfield  {author} {\bibinfo {author} {\bibfnamefont {B.}~\bibnamefont
  {Braunecker}}, \bibinfo {author} {\bibfnamefont {G.~I.}\ \bibnamefont
  {Japaridze}}, \bibinfo {author} {\bibfnamefont {J.}~\bibnamefont
  {Klinovaja}}, \ and\ \bibinfo {author} {\bibfnamefont {D.}~\bibnamefont
  {Loss}},\ }\href {\doibase 10.1103/PhysRevB.82.045127} {\bibfield  {journal}
  {\bibinfo  {journal} {Phys. Rev. B}\ }\textbf {\bibinfo {volume} {82}},\
  \bibinfo {pages} {045127} (\bibinfo {year} {2010})}\BibitemShut {NoStop}%
\bibitem [{\citenamefont {Klinovaja}\ and\ \citenamefont
  {Loss}(2012)}]{PhysRevB.86.085408}%
  \BibitemOpen
  \bibfield  {author} {\bibinfo {author} {\bibfnamefont {J.}~\bibnamefont
  {Klinovaja}}\ and\ \bibinfo {author} {\bibfnamefont {D.}~\bibnamefont
  {Loss}},\ }\href {\doibase 10.1103/PhysRevB.86.085408} {\bibfield  {journal}
  {\bibinfo  {journal} {Phys. Rev. B}\ }\textbf {\bibinfo {volume} {86}},\
  \bibinfo {pages} {085408} (\bibinfo {year} {2012})}\BibitemShut {NoStop}%
\bibitem [{\citenamefont {Alicea}(2012)}]{alicea2012new}%
  \BibitemOpen
  \bibfield  {author} {\bibinfo {author} {\bibfnamefont {J.}~\bibnamefont
  {Alicea}},\ }\href@noop {} {\bibfield  {journal} {\bibinfo  {journal}
  {Reports on Progress in Physics}\ }\textbf {\bibinfo {volume} {75}},\
  \bibinfo {pages} {076501} (\bibinfo {year} {2012})}\BibitemShut {NoStop}%
\bibitem [{\citenamefont {Klinovaja}\ \emph {et~al.}(2012)\citenamefont
  {Klinovaja}, \citenamefont {Stano},\ and\ \citenamefont
  {Loss}}]{PhysRevLett.109.236801}%
  \BibitemOpen
  \bibfield  {author} {\bibinfo {author} {\bibfnamefont {J.}~\bibnamefont
  {Klinovaja}}, \bibinfo {author} {\bibfnamefont {P.}~\bibnamefont {Stano}}, \
  and\ \bibinfo {author} {\bibfnamefont {D.}~\bibnamefont {Loss}},\ }\href
  {\doibase 10.1103/PhysRevLett.109.236801} {\bibfield  {journal} {\bibinfo
  {journal} {Phys. Rev. Lett.}\ }\textbf {\bibinfo {volume} {109}},\ \bibinfo
  {pages} {236801} (\bibinfo {year} {2012})}\BibitemShut {NoStop}%
\bibitem [{\citenamefont {Rainis}\ \emph {et~al.}(2013)\citenamefont {Rainis},
  \citenamefont {Trifunovic}, \citenamefont {Klinovaja},\ and\ \citenamefont
  {Loss}}]{PhysRevB.87.024515}%
  \BibitemOpen
  \bibfield  {author} {\bibinfo {author} {\bibfnamefont {D.}~\bibnamefont
  {Rainis}}, \bibinfo {author} {\bibfnamefont {L.}~\bibnamefont {Trifunovic}},
  \bibinfo {author} {\bibfnamefont {J.}~\bibnamefont {Klinovaja}}, \ and\
  \bibinfo {author} {\bibfnamefont {D.}~\bibnamefont {Loss}},\ }\href {\doibase
  10.1103/PhysRevB.87.024515} {\bibfield  {journal} {\bibinfo  {journal} {Phys.
  Rev. B}\ }\textbf {\bibinfo {volume} {87}},\ \bibinfo {pages} {024515}
  (\bibinfo {year} {2013})}\BibitemShut {NoStop}%
\bibitem [{\citenamefont {Jackiw}\ and\ \citenamefont
  {Rebbi}(1976)}]{PhysRevD.13.3398}%
  \BibitemOpen
  \bibfield  {author} {\bibinfo {author} {\bibfnamefont {R.}~\bibnamefont
  {Jackiw}}\ and\ \bibinfo {author} {\bibfnamefont {C.}~\bibnamefont {Rebbi}},\
  }\href {\doibase 10.1103/PhysRevD.13.3398} {\bibfield  {journal} {\bibinfo
  {journal} {Phys. Rev. D}\ }\textbf {\bibinfo {volume} {13}},\ \bibinfo
  {pages} {3398} (\bibinfo {year} {1976})}\BibitemShut {NoStop}%
\bibitem [{\citenamefont {Su}\ \emph {et~al.}(1979)\citenamefont {Su},
  \citenamefont {Schrieffer},\ and\ \citenamefont
  {Heeger}}]{PhysRevLett.42.1698}%
  \BibitemOpen
  \bibfield  {author} {\bibinfo {author} {\bibfnamefont {W.~P.}\ \bibnamefont
  {Su}}, \bibinfo {author} {\bibfnamefont {J.~R.}\ \bibnamefont {Schrieffer}},
  \ and\ \bibinfo {author} {\bibfnamefont {A.~J.}\ \bibnamefont {Heeger}},\
  }\href {\doibase 10.1103/PhysRevLett.42.1698} {\bibfield  {journal} {\bibinfo
   {journal} {Phys. Rev. Lett.}\ }\textbf {\bibinfo {volume} {42}},\ \bibinfo
  {pages} {1698} (\bibinfo {year} {1979})}\BibitemShut {NoStop}%
\bibitem [{\citenamefont {Kivelson}\ and\ \citenamefont
  {Schrieffer}(1982)}]{PhysRevB.25.6447}%
  \BibitemOpen
  \bibfield  {author} {\bibinfo {author} {\bibfnamefont {S.}~\bibnamefont
  {Kivelson}}\ and\ \bibinfo {author} {\bibfnamefont {J.~R.}\ \bibnamefont
  {Schrieffer}},\ }\href {\doibase 10.1103/PhysRevB.25.6447} {\bibfield
  {journal} {\bibinfo  {journal} {Phys. Rev. B}\ }\textbf {\bibinfo {volume}
  {25}},\ \bibinfo {pages} {6447} (\bibinfo {year} {1982})}\BibitemShut
  {NoStop}%
\bibitem [{\citenamefont {Klinovaja}\ and\ \citenamefont
  {Loss}(2013{\natexlab{b}})}]{PhysRevLett.110.126402}%
  \BibitemOpen
  \bibfield  {author} {\bibinfo {author} {\bibfnamefont {J.}~\bibnamefont
  {Klinovaja}}\ and\ \bibinfo {author} {\bibfnamefont {D.}~\bibnamefont
  {Loss}},\ }\href {\doibase 10.1103/PhysRevLett.110.126402} {\bibfield
  {journal} {\bibinfo  {journal} {Phys. Rev. Lett.}\ }\textbf {\bibinfo
  {volume} {110}},\ \bibinfo {pages} {126402} (\bibinfo {year}
  {2013}{\natexlab{b}})}\BibitemShut {NoStop}%
\bibitem [{Note1()}]{Note1}%
  \BibitemOpen
  \bibinfo {note} {In the Supplemental Material we consider higher filling
  factors and the dependence on the strip width.}\BibitemShut {Stop}%
\bibitem [{\citenamefont {Wan}\ \emph {et~al.}(2011)\citenamefont {Wan},
  \citenamefont {Turner}, \citenamefont {Vishwanath},\ and\ \citenamefont
  {Savrasov}}]{PhysRevB.83.205101}%
  \BibitemOpen
  \bibfield  {author} {\bibinfo {author} {\bibfnamefont {X.}~\bibnamefont
  {Wan}}, \bibinfo {author} {\bibfnamefont {A.~M.}\ \bibnamefont {Turner}},
  \bibinfo {author} {\bibfnamefont {A.}~\bibnamefont {Vishwanath}}, \ and\
  \bibinfo {author} {\bibfnamefont {S.~Y.}\ \bibnamefont {Savrasov}},\ }\href
  {\doibase 10.1103/PhysRevB.83.205101} {\bibfield  {journal} {\bibinfo
  {journal} {Phys. Rev. B}\ }\textbf {\bibinfo {volume} {83}},\ \bibinfo
  {pages} {205101} (\bibinfo {year} {2011})}\BibitemShut {NoStop}%
\bibitem [{\citenamefont {Yang}\ \emph {et~al.}(2011)\citenamefont {Yang},
  \citenamefont {Lu},\ and\ \citenamefont {Ran}}]{PhysRevB.84.075129}%
  \BibitemOpen
  \bibfield  {author} {\bibinfo {author} {\bibfnamefont {K.-Y.}\ \bibnamefont
  {Yang}}, \bibinfo {author} {\bibfnamefont {Y.-M.}\ \bibnamefont {Lu}}, \ and\
  \bibinfo {author} {\bibfnamefont {Y.}~\bibnamefont {Ran}},\ }\href {\doibase
  10.1103/PhysRevB.84.075129} {\bibfield  {journal} {\bibinfo  {journal} {Phys.
  Rev. B}\ }\textbf {\bibinfo {volume} {84}},\ \bibinfo {pages} {075129}
  (\bibinfo {year} {2011})}\BibitemShut {NoStop}%
\bibitem [{\citenamefont {Burkov}\ and\ \citenamefont
  {Balents}(2011)}]{PhysRevLett.107.127205}%
  \BibitemOpen
  \bibfield  {author} {\bibinfo {author} {\bibfnamefont {A.~A.}\ \bibnamefont
  {Burkov}}\ and\ \bibinfo {author} {\bibfnamefont {L.}~\bibnamefont
  {Balents}},\ }\href {\doibase 10.1103/PhysRevLett.107.127205} {\bibfield
  {journal} {\bibinfo  {journal} {Phys. Rev. Lett.}\ }\textbf {\bibinfo
  {volume} {107}},\ \bibinfo {pages} {127205} (\bibinfo {year}
  {2011})}\BibitemShut {NoStop}%
\bibitem [{\citenamefont {Xu}\ \emph {et~al.}(2011)\citenamefont {Xu},
  \citenamefont {Weng}, \citenamefont {Wang}, \citenamefont {Dai},\ and\
  \citenamefont {Fang}}]{PhysRevLett.107.186806}%
  \BibitemOpen
  \bibfield  {author} {\bibinfo {author} {\bibfnamefont {G.}~\bibnamefont
  {Xu}}, \bibinfo {author} {\bibfnamefont {H.}~\bibnamefont {Weng}}, \bibinfo
  {author} {\bibfnamefont {Z.}~\bibnamefont {Wang}}, \bibinfo {author}
  {\bibfnamefont {X.}~\bibnamefont {Dai}}, \ and\ \bibinfo {author}
  {\bibfnamefont {Z.}~\bibnamefont {Fang}},\ }\href {\doibase
  10.1103/PhysRevLett.107.186806} {\bibfield  {journal} {\bibinfo  {journal}
  {Phys. Rev. Lett.}\ }\textbf {\bibinfo {volume} {107}},\ \bibinfo {pages}
  {186806} (\bibinfo {year} {2011})}\BibitemShut {NoStop}%
\bibitem [{\citenamefont {Zyuzin}\ and\ \citenamefont
  {Burkov}(2012)}]{PhysRevB.86.115133}%
  \BibitemOpen
  \bibfield  {author} {\bibinfo {author} {\bibfnamefont {A.~A.}\ \bibnamefont
  {Zyuzin}}\ and\ \bibinfo {author} {\bibfnamefont {A.~A.}\ \bibnamefont
  {Burkov}},\ }\href {\doibase 10.1103/PhysRevB.86.115133} {\bibfield
  {journal} {\bibinfo  {journal} {Phys. Rev. B}\ }\textbf {\bibinfo {volume}
  {86}},\ \bibinfo {pages} {115133} (\bibinfo {year} {2012})}\BibitemShut
  {NoStop}%
\bibitem [{\citenamefont {Baum}\ \emph {et~al.}(2015)\citenamefont {Baum},
  \citenamefont {Posske}, \citenamefont {Fulga}, \citenamefont {Trauzettel},\
  and\ \citenamefont {Stern}}]{PhysRevLett.114.136801}%
  \BibitemOpen
  \bibfield  {author} {\bibinfo {author} {\bibfnamefont {Y.}~\bibnamefont
  {Baum}}, \bibinfo {author} {\bibfnamefont {T.}~\bibnamefont {Posske}},
  \bibinfo {author} {\bibfnamefont {I.~C.}\ \bibnamefont {Fulga}}, \bibinfo
  {author} {\bibfnamefont {B.}~\bibnamefont {Trauzettel}}, \ and\ \bibinfo
  {author} {\bibfnamefont {A.}~\bibnamefont {Stern}},\ }\href {\doibase
  10.1103/PhysRevLett.114.136801} {\bibfield  {journal} {\bibinfo  {journal}
  {Phys. Rev. Lett.}\ }\textbf {\bibinfo {volume} {114}},\ \bibinfo {pages}
  {136801} (\bibinfo {year} {2015})}\BibitemShut {NoStop}%
\bibitem [{Note2()}]{Note2}%
  \BibitemOpen
  \bibinfo {note} {One can set the system parameters and disorder strength in
  such a way that only QHE edge states will survive.}\BibitemShut {Stop}%
\bibitem [{\citenamefont {Lewenstein}\ \emph {et~al.}(2007)\citenamefont
  {Lewenstein}, \citenamefont {Sanpera}, \citenamefont {Ahufinger},
  \citenamefont {Damski}, \citenamefont {Sen},\ and\ \citenamefont
  {Sen}}]{lewenstein2007ultracold}%
  \BibitemOpen
  \bibfield  {author} {\bibinfo {author} {\bibfnamefont {M.}~\bibnamefont
  {Lewenstein}}, \bibinfo {author} {\bibfnamefont {A.}~\bibnamefont {Sanpera}},
  \bibinfo {author} {\bibfnamefont {V.}~\bibnamefont {Ahufinger}}, \bibinfo
  {author} {\bibfnamefont {B.}~\bibnamefont {Damski}}, \bibinfo {author}
  {\bibfnamefont {A.}~\bibnamefont {Sen}}, \ and\ \bibinfo {author}
  {\bibfnamefont {U.}~\bibnamefont {Sen}},\ }\href@noop {} {\bibfield
  {journal} {\bibinfo  {journal} {Advances in Physics}\ }\textbf {\bibinfo
  {volume} {56}},\ \bibinfo {pages} {243} (\bibinfo {year} {2007})}\BibitemShut
  {NoStop}%
\bibitem [{\citenamefont {Aidelsburger}\ \emph {et~al.}(2013)\citenamefont
  {Aidelsburger}, \citenamefont {Atala}, \citenamefont {Lohse}, \citenamefont
  {Barreiro}, \citenamefont {Paredes},\ and\ \citenamefont
  {Bloch}}]{PhysRevLett.111.185301}%
  \BibitemOpen
  \bibfield  {author} {\bibinfo {author} {\bibfnamefont {M.}~\bibnamefont
  {Aidelsburger}}, \bibinfo {author} {\bibfnamefont {M.}~\bibnamefont {Atala}},
  \bibinfo {author} {\bibfnamefont {M.}~\bibnamefont {Lohse}}, \bibinfo
  {author} {\bibfnamefont {J.~T.}\ \bibnamefont {Barreiro}}, \bibinfo {author}
  {\bibfnamefont {B.}~\bibnamefont {Paredes}}, \ and\ \bibinfo {author}
  {\bibfnamefont {I.}~\bibnamefont {Bloch}},\ }\href {\doibase
  10.1103/PhysRevLett.111.185301} {\bibfield  {journal} {\bibinfo  {journal}
  {Phys. Rev. Lett.}\ }\textbf {\bibinfo {volume} {111}},\ \bibinfo {pages}
  {185301} (\bibinfo {year} {2013})}\BibitemShut {NoStop}%
\bibitem [{\citenamefont {Aidelsburger}\ \emph {et~al.}(2015)\citenamefont
  {Aidelsburger}, \citenamefont {Lohse}, \citenamefont {Schweizer},
  \citenamefont {Atala}, \citenamefont {Barreiro}, \citenamefont {Nascimbene},
  \citenamefont {Cooper}, \citenamefont {Bloch},\ and\ \citenamefont
  {Goldman}}]{aidelsburger2015measuring}%
  \BibitemOpen
  \bibfield  {author} {\bibinfo {author} {\bibfnamefont {M.}~\bibnamefont
  {Aidelsburger}}, \bibinfo {author} {\bibfnamefont {M.}~\bibnamefont {Lohse}},
  \bibinfo {author} {\bibfnamefont {C.}~\bibnamefont {Schweizer}}, \bibinfo
  {author} {\bibfnamefont {M.}~\bibnamefont {Atala}}, \bibinfo {author}
  {\bibfnamefont {J.~T.}\ \bibnamefont {Barreiro}}, \bibinfo {author}
  {\bibfnamefont {S.}~\bibnamefont {Nascimbene}}, \bibinfo {author}
  {\bibfnamefont {N.}~\bibnamefont {Cooper}}, \bibinfo {author} {\bibfnamefont
  {I.}~\bibnamefont {Bloch}}, \ and\ \bibinfo {author} {\bibfnamefont
  {N.}~\bibnamefont {Goldman}},\ }\href@noop {} {\bibfield  {journal} {\bibinfo
   {journal} {Nature Physics}\ }\textbf {\bibinfo {volume} {11}},\ \bibinfo
  {pages} {162} (\bibinfo {year} {2015})}\BibitemShut {NoStop}%
\bibitem [{\citenamefont {Kraus}\ \emph {et~al.}(2012)\citenamefont {Kraus},
  \citenamefont {Lahini}, \citenamefont {Ringel}, \citenamefont {Verbin},\ and\
  \citenamefont {Zilberberg}}]{PhysRevLett.109.106402}%
  \BibitemOpen
  \bibfield  {author} {\bibinfo {author} {\bibfnamefont {Y.~E.}\ \bibnamefont
  {Kraus}}, \bibinfo {author} {\bibfnamefont {Y.}~\bibnamefont {Lahini}},
  \bibinfo {author} {\bibfnamefont {Z.}~\bibnamefont {Ringel}}, \bibinfo
  {author} {\bibfnamefont {M.}~\bibnamefont {Verbin}}, \ and\ \bibinfo {author}
  {\bibfnamefont {O.}~\bibnamefont {Zilberberg}},\ }\href {\doibase
  10.1103/PhysRevLett.109.106402} {\bibfield  {journal} {\bibinfo  {journal}
  {Phys. Rev. Lett.}\ }\textbf {\bibinfo {volume} {109}},\ \bibinfo {pages}
  {106402} (\bibinfo {year} {2012})}\BibitemShut {NoStop}%
\end{thebibliography}%

\appendix
\section{Dependence on  the width $W$ of the strip.}
\begin{figure*}[]
\centering
\includegraphics[width=18.3cm]{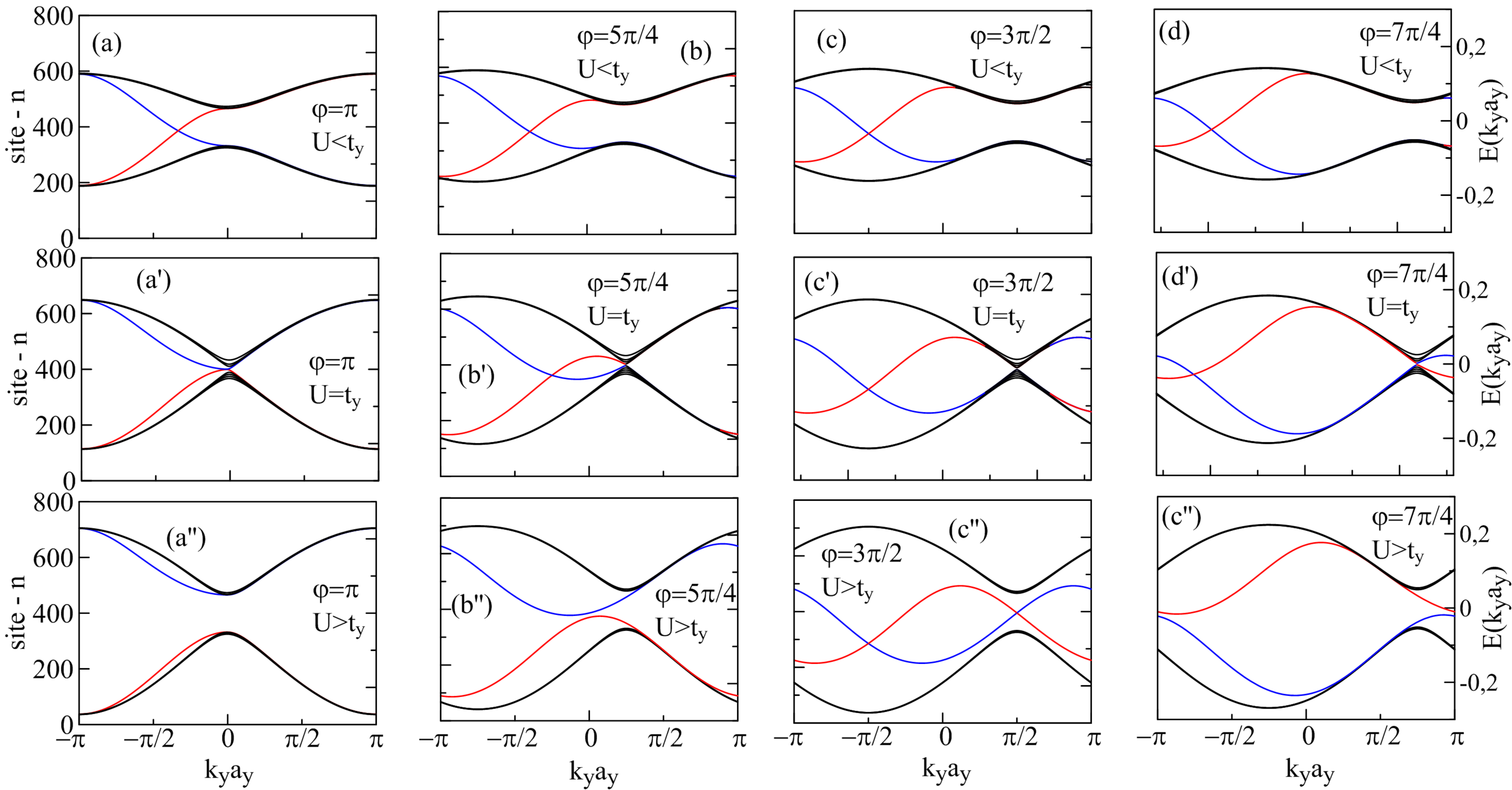}
\caption{\label{fig:nan} The energy spectrum $E(k_y)$ near the band gap for the system (a-d) in the topological (QHE dominated) phase ($|t_y|>|U|=0.05t$), (a'-d') at the phase transition point($|t_y|=|U|=0.1t$), and (a''-d'') in the non-topological (CDW dominated) phase ($|t_y|<|U|=0.15t$). The panels (a), (b), (c), and (d) refer to the phase of CDW  $\varphi=\pi,\ 5\pi/4,\ 3\pi/2,\ 7\pi/4$, respectively. The spectrum of localized edged state is found numerically (dashed line) and analytically (red line). The color map represents $|\Psi(n,k_y)|^2$ for the edge state. The position, site $n$, (energy) is marked on the left (right) axis. The found behavior of edge states is, generally, the same as described in the main text, confirming that proposed phases do not depend qualitatively on the system width.}
\label{fig:E_WF_2E}
\end{figure*}

\begin{figure}[]
\centering
\includegraphics[width=8.cm]{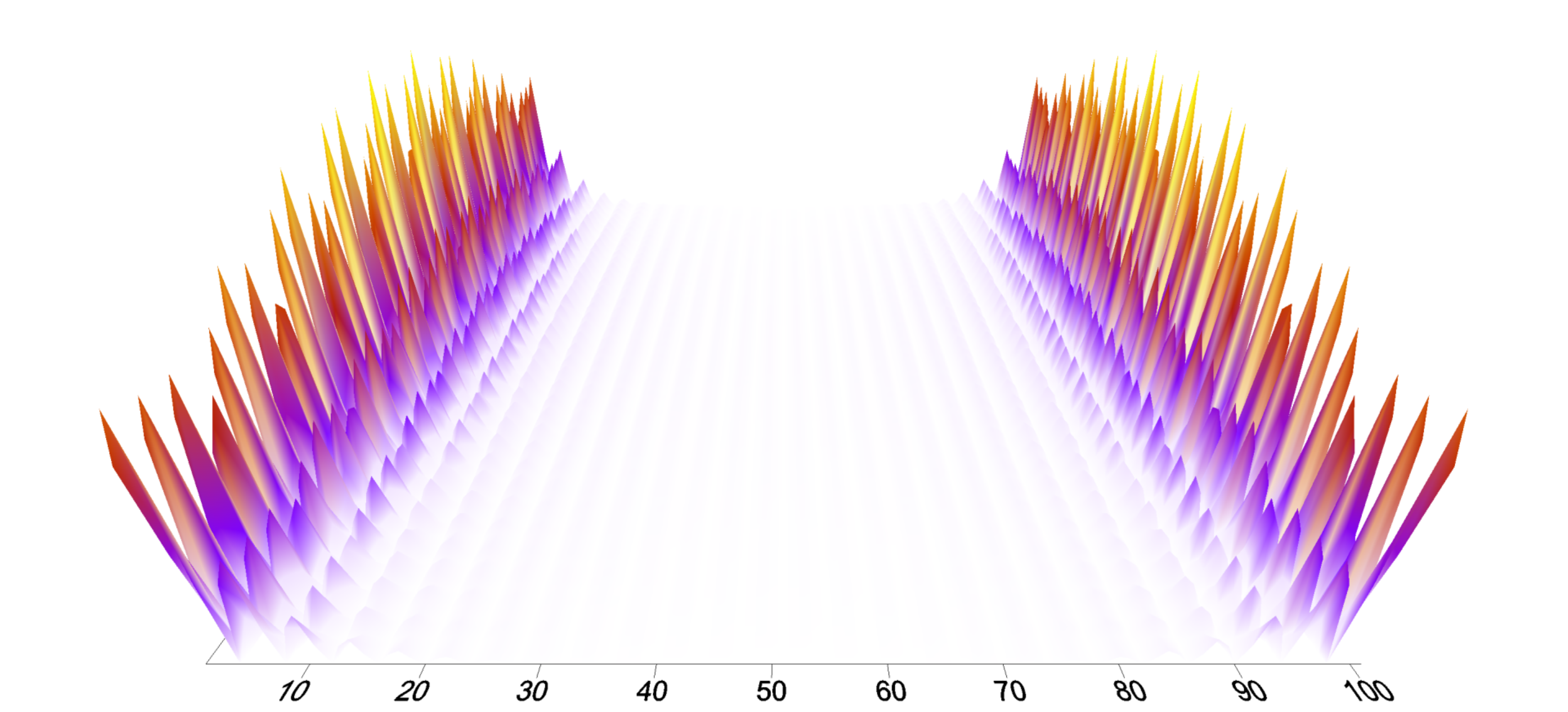}
\caption{ The wavefunction $|\Psi(n,m,E\approx 0)|^2$ for the zero edge states inside the gap in the non-topological phase $|t_y|<|U|=0.15t$ and $\varphi=3/2\pi$ where both edge states are present, see Fig. \ref{fig:E_WF_2E}.	}
\label{fig:extended_FF}
\end{figure}

In the numerical results presented in the main part, the number of lattice sites is chosen to cover an integer number of Fermi wavelengths inside the strip. As a result, the discretized CDW potential has an inversion symmetry point for $\varphi=-\pi, 0, \pi$.  However, one can choose the number of sites such that the length of the wire ({\it i.e.}, the width $W$ of the strip) corresponds to a half integer number of Fermi wavelengths. In this case the CDW phase shift $\varphi$ at the right end of the strip is different from those in the main text, see Fig. \ref{fig:E_WF_2E}. As expected, the left edge modes are not affected by  this new choice, but the dispersion of the right edge mode is changed. For example, the discretized CDW potential has now an inversion symmetry point for the  phase value $\varphi=3\pi/2$. Interestingly, in the non-topological phase, $U>t_y$, one can find again non-chiral edge states that do not touch the bulk modes. The 2D wave functions $|\Psi(n,m,\varepsilon\approx 0)|^2$ for  both edge states present in the non-topological regime are depicted in the Fig. \ref{fig:extended_FF}.

\section{Filling factor $\nu=2$.}
One can also observe phase transitions similar to the ones descibed in the main text for $\nu=1$  for the filing factor $\nu=2$, which corresponds to an indirect resonant magnetic field with $\phi=k_Fx$~\cite{PhysRevLett.111.196401,FQHE_CDW_Jelena}. In this case, the size of the gap opened by the resonant scattering is smaller ($\Delta~\approx~2t_y^2/\mu$) than for the $\nu=1$ case and the phase transitions occurs for correspondingly smaller $U_0$. In the topological phase ($U_0=0.01$, $\varphi=0, -\pi/2$), there are always two chiral edge states present at each edge, see Figs. 3 (a)-(b). Interestingly, in the non-topological phase ($U_0=0.05$, $\varphi=-\pi/2$), there can be two or four modes present at the same edge, see Fig. \ref{fig:E_WF_phi_0}(b').

\begin{figure}[]
\centering
\includegraphics[width=8.6cm]{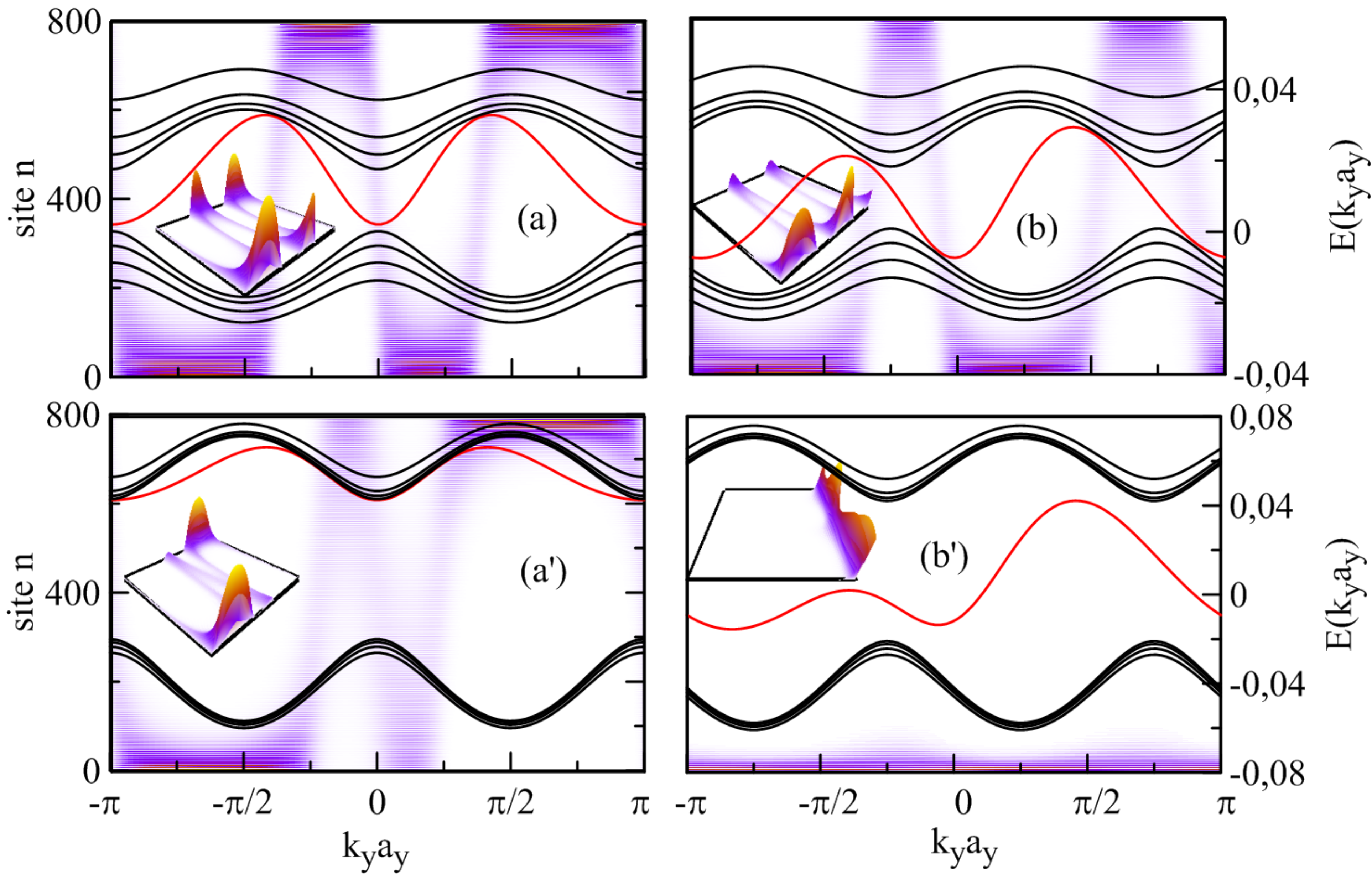}
\caption{\label{fig:2D_CDW_DIS} The same as in Fig. 1 but for the filling factor $\nu=2$ (resulting in $\phi=k_Fa_x$). In the topological phase we can see two chiral edge states present at each edge. In the non-topological phase, depending on the chemical potential, there could be two or four non-chiral edge states,  see the panel (b'). }
\label{fig:E_WF_phi_0}
\end{figure}

\newpage

\bigskip 

\newpage

\bigskip

\end{document}